\documentclass[aps,prb,twocolumn,groupedaddress,showpacs]{revtex4}

\usepackage{graphicx}
\usepackage{dcolumn}
\usepackage{bm}
\usepackage{amsmath}

\newcommand{\ket}[1]{\ensuremath{\left| #1 \right \rangle }} 
\newcommand{\bra}[1]{\ensuremath{\left \langle #1 \right |}}
\newcommand{\kPhi}{\ensuremath{\left| \Phi \right \rangle}}
\newcommand{\kPhin}{\ensuremath{\left| \Phi_0 \right \rangle}}

\newcommand{\s}{\ensuremath{\sigma}}
\newcommand{\te}{\ensuremath{\mathrm t}}
\newcommand{\re}{\ensuremath{\mathrm r}}

\begin{document}


\title{Landau--Gutzwiller quasi-particles}

\author {J\"org B\"unemann}
\affiliation{Oxford University, Physical and Theoretical Chemistry Laboratory, 
South Parks Road, Oxford OX1 3QZ, United Kingdom}
\author {Florian Gebhard}
\affiliation{Fachbereich Physik, Philipps--Universit\"at Marburg, 
D--35032 Marburg, Germany}
\author {R\"udiger Thul}
\affiliation{Abteilung Theorie, Hahn-Meitner-Institut Berlin,
D--14109 Berlin, Germany}


\begin{abstract}%
We define Landau quasi-particles within the Gutzwiller variational theory,
and derive their dispersion relation for
general multi-band Hubbard models in the limit of large spatial 
dimensions~$D$. Thereby 
we reproduce our previous calculations which were based on  
a phenomenological effective single-particle Hamiltonian.
For the one-band Hubbard model we calculate the first-order corrections 
in $1/D$ and find that the corrections to the quasi-particle dispersions
are small in three dimensions. They may be largely absorbed in
a rescaling of the total band width, 
unless the system is close to half band filling.
Therefore, the Gutzwiller theory in the limit of large
dimensions provides quasi-particle bands which are
suitable for a comparison with real, three-dimensional Fermi liquids.
\end{abstract}


\maketitle

\section{Introduction}
\label{sec1}

The calculation of the band structure of metals 
and insulators is a central task in solid-state theory. 
A commonly accepted method for this purpose 
is the 
density-functional theory (DFT) 
which provides surprisingly accurate results 
for the band structure of many materials~\cite{Moruzzi}. 
Furthermore, the DFT is an  `ab-initio' theory, i.e., 
it starts from the full Hamiltonian 
of a real system and does not require the introduction 
of any simplified models. 
The only competing `ab-initio' theory was
Hartree-Fock theory, which has shown very many shortcomings as
compared to DFT, such as gross overestimation of band widths and
band gaps.

However, from a theoretical point of view, the 
success of the DFT for band structures is rather astonishing 
because this theory is a generic approach 
to ground-state properties only. 
All results on energy bands are extracted from 
auxiliary one-particle dispersions 
which have no physical meaning at the outset. 
Indeed, some shortcomings of the DFT energy bands have become
evident very early, in particular
the underestimation of the fundamental gap in semiconductors.
In semiconductors and insulators, the so-called GW approximation
to the one-particle Green function
has been put forward~\cite{Louie}. There, the single-particle self energy
is calculated using a Green function based on the DFT
wave functions and the screened Coulomb interaction.
It turns out that the GW quasi-particle bands are more or less rigidly shifted
against the DFT bands so that
the band gap results of GW~calculations for semiconductors and insulators
agree much better with experiment. 

For materials with strong electron-electron interactions, the DFT results
have not been too convincing, in particular for magnetic insulators and
other strongly correlated electron systems. For the iron group metals 
the discrepancies of DFT results to experimental data, e.g.,
angle-resolved photo-emission spectroscopy (ARPES), increase
towards the end of the series, i.e., towards nickel;
for a detailed discussion on the discrepancies between DFT results
and experimental data on nickel, see Refs.~\onlinecite{gutzjubel,wandlitz}.
For the iron group metals, GW calculations did not yield
significant improvements over the DFT results; for nickel, see
Ref.~\onlinecite{Ferdi}.

A proper description of solids with strong 
Coulomb interactions requires true
many-particle approaches. In the past, the notorious difficulties
of many-particle systems have restricted
such theories to the study of rather simplified model systems, e.g.,
the one-band Hubbard model. Therefore,
a comparison with experiments on real materials could hardly be performed.
Only recently, new non-perturbative many-particle methods
have become available which have made possible the investigation 
of more realistic many-particle models; see, for example,
Refs.~\onlinecite{gutzjubel,wandlitz,Pruschke,Vollhardt,PRB98}.
In Ref.~\onlinecite{PRB98} we
introduced a class of Gutzwiller variational 
wave functions which allow us to study general multi-band Hubbard models. 
Expectation values with these correlated electron states are
evaluated exactly in the limit of large spatial dimensions, $D\to\infty$.
When applied to nickel, the remaining minimization problem
is numerically non-trivial because of the large number of 
variational parameters; first results 
are reported in Refs.~\onlinecite{gutzjubel,wandlitz}. 

The Gutzwiller variational theory provides an approximate picture of the 
ground state but, in principle, 
it lacks any information about excited states. 
This drawback can be overcome in two ways.
First, if we take for granted that the variational 
ground state is at least qualitatively close to the true ground state, 
we may use the variational state as a starting point 
for the variational calculation of excited states.
In Ref.~\onlinecite{spinpaper} we have used this idea  
to determine the spin-wave dispersion in ferromagnetic multi-band
Hubbard models. We have successfully reproduced the 
experimental observation that the low-energy spin excitations 
in itinerant ferromagnets are very similar to 
those of a system with localized spins. 
Second, the calculation of the variational ground-state energy in
the limit of infinite dimensions~\cite{PRB98} naturally leads to the
definition of an effective single-particle Hamiltonian
which, in some limits, can also be derived in Slave-Boson mean-field
theory~\cite{KR}. Very much in the spirit of the DFT we used
the band structure of this effective Hamiltonian for a successful
comparison with ARPES data for nickel. We have been able to resolve 
basically all of the LDA shortcomings.

Despite its success, the second approach still lacks a sound theoretical basis.
In this work we derive the (variational) quasi-particle dispersion 
referring back to Landau's original ideas on Fermi liquids.
The Gutzwiller variational state is an illustrative example 
for a Fermi-liquid ground state: the Gutzwiller many-body correlator
acts on the Fermi-gas ground state whereby energetically unfavorable
configurations are gradually reduced. In the spirit of Fermi-liquid
theory, a quasi-particle excitation is readily viewed as
a Gutzwiller-correlated single-particle excitation of the Fermi-gas
ground state. The energy of this excitation
is identical to the quasi-particle dispersion 
in our original work~\cite{PRB98}. 
Therefore, no revision of our previous  
numerical results on nickel~\cite{gutzjubel,wandlitz} is necessary. 

The evaluation of the variational energy  
in our method is exact only in the limit of 
infinite spatial dimensions. Our application
to realistic three-dimensional systems 
requires that $1/D$~corrections are well controlled.
As is known from the one-band model,
these corrections are small for ground-state properties 
such as the (variational) energy or the effective mass of the quasi-particles 
at the Fermi surface. 
In this work we will present additional results 
on~$1/D$~corrections of the quasi-particle dispersion for the one-band 
Hubbard model.

Our work is organized as follows. In Sec.~\ref{sec2} we 
summarize the basic ideas of Landau-Gutzwiller theory.
In Sect.~\ref{sec3} we discuss the variational ground state 
for multi-band Hubbard models.
In Sec.~\ref{sec4} we define Landau-Gutzwiller quasi-particles
and derive their energy dispersion.
In Sect.~\ref{sec5} we 
calculate $1/D$~corrections 
for the quasi-particle dispersion of the one-band model.       
Our presentation closes with short conclusions.
Some technical details are deferred to the appendix.

\section{Landau-Gutzwiller Theory}
\label{sec2}

In second quantization the Hamilton operator for non-interacting electrons
reads 
\begin{equation}
\label{10single}
\widehat{H}_1=\sum_{i\neq j;\s,\s'}
t_{i,j}^{\s,\s'}
\widehat{c}_{i;\s}^{\, +}
\widehat{c}_{j;\s'}^{\vphantom{\, +}}
+ \sum_{i;\s}
\widetilde{\epsilon}_{\s} \widehat{c}_{i;\s}^{\, +}
\widehat{c}_{i;\s}^{\vphantom{\, +}}
 \; .
\end{equation}
Here, $\widehat{c}_{i;\s}^{\, +}$ creates an electron with combined
spin-orbit index~$\s=1,\ldots ,2N$ ($N=5$ for 3$d$~electrons) at
the lattice site~$i$ of a solid with $L$~lattice sites.
The electron density, $n=N_{\rm e}/L$, is finite in the thermodynamic limit
$N_{\rm e}\to\infty$, $L\to\infty$.

For a translationally invariant system,
as considered throughout this work,
this single-particle Hamiltonian is readily diagonalized
in momentum space. Its eigenstates are
one-particle product states~$|\Psi\rangle$. In particular,
the ground state~$|\Psi_0\rangle$ is the filled Fermi sea
where all single-particle states below the Fermi energy
are occupied. All other eigenstates can be understood as particle-hole
excitations of $|\Psi_0\rangle$. 

One essential idea behind Landau's Fermi liquid theory is the assumption that
the Fermi-gas picture remains valid qualitatively when
electron-electron interactions are switched on;
for an introduction, see, e.g., Ref.~\onlinecite{nozieres}. 
The Fermi-gas eigenstates
transform adiabatically into those of the Fermi liquid while keeping
their physical properties. For example, the momentum distribution
displays a discontinuity at the Fermi energy both in the Fermi gas 
and in the Fermi liquid. Naturally, the properties of the ground state
and of the particle-hole excitations change quantitatively.
Therefore, the excitations are called quasi-particles and 
quasi-holes in the Fermi liquid.

Gutzwiller's variational theory closely follows the idea
of an adiabatic continuity from the Fermi gas to the Fermi liquid.
Let us introduce a general class of Gutzwiller-correlated wave functions
via
\begin{equation}
\label{30}
|\Psi_{\rm G}\rangle =\widehat{P}_{\rm G}|\Phi\rangle \;.
\end{equation}
Here, $|\Phi\rangle$ is any normalized one-particle product state.
The Gutzwiller correlator
\begin{equation}
\widehat{P}_{\rm G} = \prod_i\widehat{P}_{i;{\rm G}} 
\end{equation}
is a many-body operator which suppresses those configurations which
are energetically unfavorable with respect to the electron-electron
interaction. Therefore, the (approximate) Fermi-liquid ground
state 
\begin{equation}
|\Psi^0_{\rm G}\rangle = \widehat{P}_{\rm G}|\Phi_0\rangle
\end{equation}
evolves smoothly from 
a Fermi-gas ground state $|\Phi_0\rangle$ when the electron-electron
interaction is switched on. In fact, this concept has been
used by Vollhardt~\cite{RMPVollhardt} to develop
a microscopic theory for the ground-state properties of
liquid $^{3}$He on the basis of Gutzwiller's approach.

In this work, we use Landau's idea to extend Gutzwiller's variational
approach to quasi-particle excitations.
In principle, this does not pose a big problem. Instead of
a Fermi-liquid ground state~$|\Phi_0\rangle$, we use
single-particle excitations $|\Phi\rangle$ of the Fermi gas
in~(\ref{30}) to define quasi-particle states.
In Sect.~\ref{sec4} we will give a proper mathematical definition of
a quasi-particle excitation.
Here we point out that 
all restrictions of Fermi-liquid theory apply. For example,
only the low-energy properties of metals, close to the
Fermi energy, ought to be described in this way. 
Nevertheless, experiments on metals show that well-defined 
but life-time broadened quasi-particle excitations 
can be found even for energies of about $10\, {\rm eV}$ 
below the Fermi energy. Therefore, the concept of
quasi-particles and quasi-holes remains meaningful
for those parts of the valence and conduction bands
which are relevant in solid-state physics.

\section{Variational energy}
\label{sec3}

\subsection{Multi-band Hubbard Hamiltonian and Gutzwiller variational states}

In the following we study multi-band Hubbard models where the
electron-electron interaction is purely local,
\begin{equation}
\label{10}
\widehat{H}=\widehat{H}_1 + \sum_i\widehat{H}_{i;{\rm at}} \; .
\end{equation}
Here, the atomic Hamiltonian $\widehat{H}_{i;{\rm at}}$ 
contains all possible Coulomb-interaction terms between electrons
on site~$i$, 
\begin{equation}
\label{20}
\widehat{H}_{i;{\rm at}} =
\sum_{\s_1,\s_2,\s_3,\s_4}
{\cal U}^{\s_1,\s_2;\s_3,\s_4}
\widehat{c}_{i;\s_1}^{\, +}\widehat{c}_{i;\s_2}^{\, +}
\widehat{c}_{i;\s_3}^{\vphantom{\, +}}
\widehat{c}_{i;\s_4}^{\vphantom{\, +}}\;.
\end{equation}
We assume that the eigenstates $|\Gamma\rangle_i$ of the atomic 
Hamiltonian have been determined
\begin{equation}
\widehat{H}_{i;{\rm at}} 
=
\sum_{\Gamma} E_{i;\Gamma}\widehat{m}_{i;\Gamma}
\quad , \quad  
\widehat{m}_{i;\Gamma} 
= |\Gamma\rangle_i \, {}_i\langle \Gamma | \; .
\label{22}
\end{equation}
This is possible in all cases of interest, 
at least numerically. In the following, 
the site index will often be suppressed as we are primarily
interested in translationally invariant systems.

The Gutzwiller theory allows us to study the Hamiltonian~(\ref{10}) 
with an arbitrary number of orbitals~\cite{PRB98}. 
In this work, however, we will restrict ourselves to the special case 
where non-degenerate orbitals belong to different representations 
of the respective point-symmetry group. 
For example, in cubic symmetry we allow for
only one set of $s$, $p$, $d(e_g)$ and $d(t_{2g})$~orbitals.

The Gutzwiller correlator
\begin{equation}
\widehat{P}_{{\rm G}} = \prod_{i} \widehat{P}_{i;{\rm G}} =
\prod_{i;\Gamma} \lambda_{i;\Gamma }^{\widehat{m}_{i;\Gamma}}
\end{equation}
is parameterized by $2^{2N}$ real numbers~$\lambda_{i;\Gamma }$.
For an energetically 
unfavorable atomic configuration~$|\Gamma_{i}\rangle$ 
the minimization will result in $\lambda_{i;\Gamma }<1$ whereby
its weight in $|\Phi\rangle$ is reduced.

\subsection{Extrema of the variational energy}
\label{sec3.1}

In the limit of large spatial dimensions, the expectation value of 
the Hamiltonian~(\ref{10}) in the wave function~(\ref{30})  
can be expressed in terms of the one-particle product
wave function~$|\Phi\rangle$ 
and the expectation values 
\begin{equation}
\label{40}
m_{\Gamma}=\langle \widehat{m}_{\Gamma} \rangle_{\Psi_{\rm G}}
=\frac{\langle \Psi_{\mathrm G} | \widehat{m}_{\Gamma}  | \Psi_{\mathrm G}\rangle}%
{\langle \Psi_{\mathrm G}|\Psi_{\mathrm G}\rangle} \; ;
\end{equation}   
for all details, see Ref.~\onlinecite{PRB98}. 
After a lengthy calculation
one obtains the following variational energy  
\begin{subequations}
\label{43}
\begin{eqnarray}  
\label{43a}
E^{\rm var} &=&
\langle \widehat{H} \rangle_{\Psi_{\rm G}} =
\sum_{\s,\s^\prime}\sum_{{\mathbf k}}S_{\s,\s^\prime}({\mathbf k}) 
\langle \widehat{c}_{{\mathbf k};\s}^{\, +}
\widehat{c}_{{\mathbf k};\s'}^{\vphantom{\, +}} \rangle_{\Phi} 
\nonumber \\
&& +L\sum_{\Gamma }E_{\Gamma}m_{\Gamma }\;,  \\ 
S_{\s,\s^\prime}({\mathbf k}) &=& 
\sqrt{q_{\s}} \sqrt{q_{\s^\prime}} \epsilon_{\s,\s^\prime}({\mathbf k}) +   
\delta_{\s,\s^\prime} \widetilde{\epsilon}_{\s} \, , 
\label{43b}
\end{eqnarray}\end{subequations}%
where
\begin{equation}    
\epsilon_{\s,\s^\prime}({\mathbf k}) = \frac{1}{L} \sum_{l\neq m}  
t_{l,m}^{\s,\s^\prime} 
\exp\left(-{\rm i} {{\mathbf k}({\mathbf R}_l-{\mathbf R}_m)}\right)
 \; .
\end{equation}
The calculation only requires $|\Phi\rangle$ to be
a one-particle product wave state; $|\Phi\rangle$ need not be
a filled Fermi sea.

For a given set of Coulomb parameters in~(\ref{20}) 
the renormalization factors 
\begin{equation}
\label{47}
q_{\s}= q_{\s}(n^{0}_{\s},m_{\Gamma})
\end{equation} 
only depend on the local spin-orbital densities 
\begin{equation}
n^{0}_{\s}=\langle \widehat{n}_{\s}  \rangle_{\Phi}
\end{equation} 
and the variational parameters $m_{\Gamma}$ for states $\ket{\Gamma}$ with 
more than one electron.
An explicit expression for~(\ref{47}) has been given in
Ref.~\onlinecite{PRB98}, 
but it is not needed for our further considerations.
Note that for our symmetry-restricted orbital basis 
\begin{equation}
n^{0}_{\s}=n_{\s}=\langle \widehat{n}_{\s}  \rangle_{\Psi_{\rm G}}
\end{equation}
holds. 

$E^{\rm var}$ in~(\ref{43}) depends on the variational parameters 
$m_{\Gamma}$, the local densities $n_{\s}$, 
and the wave function $\kPhi$. 
However, the constraints
\begin{equation}
\label{150}
n_{\s}
= \langle\Phi|\widehat{n}_{\s} |\Phi\rangle
%
\quad ; \quad 
n
=\sum_{\s}n_{\s} 
\end{equation}
have to be respected during the minimization
as we work for fixed~$n$ in the sub-space
of normalized one-particle product states,
$\langle\Phi|\Phi\rangle=1$.
We introduce Lagrange parameters 
$E_{\rm SP}$, $\lambda_{\s}$, and $\Lambda$  
for these constraints which leads to the energy functional
\begin{eqnarray}
E_{\rm c} [\Phi, m_{\Gamma},n_{\s},\lambda_{\s},\Lambda]&=&
E^{\rm var}[\kPhi,m_{\Gamma},n_{\s}]
\nonumber\\
&& -L \sum_{\s} \lambda_{\s} 
\big[ n_{\s} - \bra{\Phi}  \widehat{n}_{\s}\ket{\Phi} \big] 
\nonumber 
\\ 
&&-L\Lambda(n-\sum_{\s}n_{\s})\label{160}
\\
&&+E_{\rm SP}(1-\langle\Phi|\Phi\rangle)\;.
\nonumber
\end{eqnarray}
$E_{\mathrm c} $ has now to be 
minimized with respect to all 
quantities $\kPhi$, $E_{\rm SP}$, $m_{\Gamma}$, $n_{\s}$, $\lambda_{\s}$,
and $\Lambda$ independently. 

First, we use the condition that~(\ref{160}) is extremal with 
respect to $\kPhi$ and $E_{\rm SP}$.  
This gives us the following effective Schr\"odinger equation 
which has to be solved in the sub-space of normalized states~$\kPhi$
\begin{equation}
\label{169}
\widehat{H}^{\, \mathrm{eff}}\kPhi=E_{\rm SP}\kPhi
\end{equation}
with
\begin{equation}
\label{170}
\widehat{H}^{\, \mathrm{eff}}=\sum_{\s,\s^\prime}
\sum_{{\mathbf k}}(S_{\s,\s^\prime}({\mathbf k})+
\delta_{\s,\s^\prime}\lambda_{\s}) \widehat{c}_{{\mathbf k};\s}^{\, +}
\widehat{c}_{{\mathbf k};\s'}^{\vphantom{\, +}}\; .
\end{equation}
The effective one-particle 
Hamiltonian $\widehat{H}^{\, \mathrm{eff}}$ can be diagonalized,
\begin{equation}
\label{172}
\widehat{H}^{\, \mathrm{eff}}=  
\sum_{{\mathbf k}, \re}E({\mathbf k}, \re) 
\widehat{h}_{{\mathbf k}, \re }^{\, +} 
\widehat{h}_{{\mathbf k}, \re }^{\vphantom{\, +}} 
\end{equation}
 by introducing  proper creation and annihilation operators
\begin{equation}
\widehat{h}_{{\mathbf k}, \re }^{\, +}
:=
\sum_{\s} F_{{\mathbf k},\s,\re }^{\vphantom{\ast}} 
\widehat{c}_{{\mathbf k},\s }^{\, +} 
\quad ; \quad 
%
\widehat{h}_{{\mathbf k},\re }^{\vphantom{\, +}}
:=
\sum_{\s} F_{{\mathbf k},\s,\re }^{\ast} 
\widehat{c}_{{\mathbf k},\s }^{\vphantom{\, +}}
\; .
\label{180}
\end{equation}
Note that the amplitudes $F_{{\mathbf k},\s,\re}^{\vphantom{\ast}}$, 
the energies $E({\mathbf k}, \re)$, 
and the operators $\widehat{h}_{{\mathbf k}, \re }^{\, +}$,
$\widehat{h}_{{\mathbf k}, \re }^{\vphantom{\, +}}$ 
still depend on the parameters $m_{\Gamma}$, $n_{\s}$, and $\lambda_{\s}$. 

Solving the eigenvalue equation~(\ref{169}) is only a necessary 
but not a sufficient condition 
for a state $\kPhi$ to minimize 
the original energy expression~(\ref{43}). 
In Sect.~\ref{sec3} we will use 
this ambiguity to define quasi-particles excitations
of the variational Fermi-liquid ground state.

\subsection{Variational Fermi-liquid ground state}

In order to obtain our variational Fermi-liquid ground state,
it appears to be the most natural choice 
to proceed with the filled Fermi sea for 
the effective Hamiltonian $\widehat{H}^{\, \mathrm{eff}}$,
\begin{equation}
\label{190}
\kPhin=\prod_{{\mathbf k}, \re; E({\mathbf k}, \re)<E_{\rm F}}
\widehat{h}_{{\mathbf k}, \re}^{\, +} 
\left | \text{vacuum} \right \rangle \;.
\end{equation}
Here, the Fermi energy $E_{\rm F}$ is determined by the condition 
\begin{equation}
\label{200}
\frac{1}{L}\sum_{{\mathbf k}, \re} 
\Theta(E_{\rm F} - E({\mathbf k}, \re)) 
=n\;.
\end{equation}
The corresponding eigenvalue $E_{\rm SP}$ becomes
\begin{equation}
\label{210}
E_{\rm SP}=\sum_{{\mathbf k}, \re}
E({\mathbf k}, \re)
\Theta(E_{\rm F} - E({\mathbf k}, \re)) 
\; .
\end{equation}
It is difficult to prove rigorously
that the state~(\ref{190}) leads to the global minimum of~(\ref{43}). 
However, $\kPhin$ is at least stable 
with respect to single-particle excitations 
and it is difficult to conceive any other state 
which is consistent with our underlying Fermi-liquid picture. 

When we insert $\kPhin$ into~(\ref{160}) we are led to the energy function
\begin{eqnarray}
\label{220}
\widetilde{E}_{\mathrm{c}} [m_{\Gamma},n_{\s},\lambda_{\s},\Lambda]
&=&E_{\rm SP}
+L\sum_{\Gamma}E_{\Gamma} m_{\Gamma}
-L\sum_{\s}\lambda_{\s} n_{\s} \nonumber \\
&& -L \Lambda(n-\sum_{\s}n_{\s}) \;.
\end{eqnarray}
In the variational ground state this expression
is extremal with respect to $m_{\Gamma}$, $n_{\s}$, $\lambda_{\s}$,
and $\Lambda$,
\begin{equation}
\label{222}
\left. \frac{\partial}{\partial x_i} 
\widetilde{E}_{\mathrm c}\right|_{\{x_j\}=\{\overline{x}_j\}}=0
\quad \text{with}\quad x_i \in \{m_{\Gamma},n_{\s},\lambda_{\s},\Lambda\}\;.
\end{equation}
The optimum values  
$\overline{m}_{\Gamma}$, $\overline{n}_{\s}$, $\overline{\lambda}_{\s}$, and
$\overline{\Lambda}$ define the optimum values for the energies
$\overline{E}({\mathbf k}, \re )$,
the amplitudes $\overline{F}_{{\mathbf k},\s,\re}^{\vphantom{\ast}}$, 
and the operators  
$\widehat{\overline{h}}{}\vphantom{\widehat{h}}_{{\mathbf k}, \re }^{\, +}$,
$\widehat{\overline{h}}{}\vphantom{\widehat{h}}_{{\mathbf k}, \re }^{\vphantom{\, +}}$.
Furthermore, we can write the variational ground-state energy as
\begin{equation}
\label{224}
E_0^{\rm var}=\widetilde{E}_{\rm c} [\overline{m}_{\Gamma},
\overline{n}_{\s},\overline{\lambda}_{\s},\overline{\Lambda}]\;.
\end{equation}

The energy~(\ref{224}) depends on the particle density~$n$ both
implicitly, mediated by the optimum values
$\overline{m}_{\Gamma}(n)$, $\overline{n}_{\s}(n)$, $\overline{\lambda}_{\s}(n)$,
and $\overline{\Lambda}(n)$, and
explicitly, due to the term $-L\overline{\Lambda}(n)n $ and the
Fermi energy  $E_{\rm F}\equiv E_{\rm F}(n)$ in $E_{\rm SP}$ of~(\ref{220}). 
Therefore, the (variational) chemical potential 
\begin{equation}
\label{114}
\mu = \frac{1}{L}\frac{d E_0^{\rm var}  }{d n}
\end{equation}
can be written as
\begin{eqnarray}
\mu&=&
\frac{1}{L}\frac{\partial E_{\rm SP}}{\partial n}-\overline{\Lambda}
\label{244}\\
&& +\frac{1}{L}\sum_{x_i \in \{m_{\Gamma},n_{\s},\lambda_{\s},\Lambda\}}
\left.\frac{\partial}{\partial x_i}\widetilde{E}_{\mathrm c}  \right|_{\{x_j\}
=\{\overline{x}_j\}}    \frac{\partial \overline{x}_i}{\partial n}\;.
\nonumber 
\end{eqnarray}
The sum in~(\ref{244}) vanishes due to~(\ref{222}) whereas the
derivative of $E_{\rm SP}$ just gives the Fermi energy $E_{\rm F}$,
\begin{equation}
\label{246}
\frac{1}{L}\frac{\partial E_{\rm SP}}{\partial n}=E_{\rm F}\;.
\end{equation}
Therefore, the variational chemical potential reads
\begin{equation}
\label{248}
\mu=E_{\rm F}-\overline{\Lambda}\;.
\end{equation}

The strategy for the numerical minimization
is not important for our analysis of Landau-Gutzwiller quasi-particles
in the rest of our work.
For further reference we give a short summary of our most efficient procedure.

First, we note that the conditions 
$(\partial \widetilde{E}_{\mathrm{c}})/(\partial \lambda_{\s})=0$ 
and $(\partial \widetilde{E}_{\mathrm c})/(\partial \Lambda)=0$ 
take us back to the original constraints~(\ref{150}),
\begin{subequations}
\begin{eqnarray}
n_{\s}&=&\frac{\partial}{\partial \lambda_{\s}} E_{\rm SP}
=\langle \widehat{n}_{\s} \rangle_{\Phi_0}\nonumber\\
&=&\sum_{{\mathbf k}, \re}|F_{{\mathbf k},\s,\re}^{\vphantom{\ast}}|^2
\Theta(E_{\rm F} - E({\mathbf k}, \re))\; , \label{230}\\
\label{232}
n&=&\sum_{\s}n_{\s}\; .
\end{eqnarray}\end{subequations}%
Therefore, we are left with two different sets of variational
parameters, the `internal' parameters $m_{\Gamma}$ and the `external'
parameters $\lambda_{\s}$. Optimizing the energy with respect to both of
these sets is time-costly, for different reasons. 
The problem with the internal parameters is their large 
number which is of the order~$2^{2N}$ ($\approx 500$ for $d$~orbitals). 
Compared to this there are
only a few, $2N$, external parameters~$\lambda_{\s}$. 
However, each modification
of one of these external parameters demands for momentum-space
integrations according to the sums in~(\ref{230}) and~(\ref{210}). We found
these integrations to be the most time-consuming part of our
numerical minimization. In principle, such integrals must also be
calculated whenever we  change the parameters $m_{\Gamma}$, because
they determine the amplitudes $F_{{\mathbf k},\s,\re}^{\vphantom{\ast}}$ 
and the energies
$E({\mathbf k},\re)$. In order to avoid this large number
of integrations we write  $E_{\rm SP}$ in~(\ref{220}) as
\begin{eqnarray}
E_{\rm SP}&=&\sum_{\s,\s^{\prime}}\sqrt{q_{\s}} \sqrt{q_{\s^{\prime}}}
\mathop{{\sum}'}_{{\mathbf k}, \re}\epsilon_{\s,\s^{\prime}}({\mathbf k}) 
F_{{\mathbf k},\s,\re}^{\ast} 
F_{{\mathbf k},\s^{\prime},\re}^{\vphantom{\ast}} 
\nonumber \\
&& +\sum_{\s}(\widetilde{\epsilon}_{\s}+\lambda_{\s} )
\mathop{{\sum}'}_{{\mathbf k}, \re}
|F_{{\mathbf k},\s,\re}^{\vphantom{\ast}} |^2 
\;,
\label{240}
\end{eqnarray}         
where the prime on the sums implies $E({\mathbf k}, \re)) < E_{\rm F}$.
Eqs.~(\ref{220}) and~(\ref{240}) show that the 
parameters $m_{\Gamma}$ enter the energy $\widetilde{E}_{\mathrm {c}}$ 
in two different ways: (i), indirectly, via the amplitudes 
$F_{{\mathbf k},\s,\re}^{\vphantom{\ast}}$
or the energies $E({\mathbf k}, \re)$ and, (ii),  
directly, via $q_{\s}$ in~(\ref{240}) and the second term in~(\ref{220}). 
This separation suggests the following numerical iteration scheme:
\begin{enumerate}
\item Start with an initial guess 
for the parameters $m_{\Gamma}$, e.g., 
their statistical values in the uncorrelated limit.
\item \label{step2}
Minimize the energy with 
respect to the parameters~$\lambda_{\s}$ while all~$m_{\Gamma}$ are fixed. 
During this minimization the constraint~(\ref{232}) must be respected.
\item  Minimize the energy with respect 
to the parameters~$m_{\Gamma}$, while the parameters~$\lambda_{\s}$, 
the amplitudes $F_{{\mathbf k},\s,\re}^{\vphantom{\ast}}$,
the energies $E({\mathbf k}, \re)$
and the wave function $\ket{\Phi_{0}}$
remain fixed during step~\ref{step3}.
\label{step3}
\item Go back to step~\ref{step2} unless the 
reduction of $E^{\rm var}$ becomes sufficiently small.
\end{enumerate}
The above procedure represents only a rough picture of 
our numerical minimization. 
For example, in practice one finds that some of the 
parameters $\lambda_{\s}$ play only  a minor role and, therefore,  
are fixed during the whole minimization. 
However, we are not going to discuss these numerical details in this work
because they depend on the specific material under investigation.

\section{Landau-Gutzwiller quasi-particles}
\label{sec4}

\subsection{Definition}
\label{sec4.1}

The Gutzwiller theory provides~$|\Psi^0_{\rm G}\rangle$,
an approximate description of the true
many-body ground state. In order to extend the variational
description to quasi-particle excitations, we closely follow Landau's
ideas. 
We seek creation and annihilation operators 
$\widehat{e}^{\, +}_{{\mathbf p},t}$ and 
$\widehat{v}_{{\mathbf p},t}^{\vphantom{\, +}}$ 
which must obey the same Fermi-Dirac distribution  
around the Fermi surface as uncorrelated electrons, i.e., we postulate
\begin{equation}
 \label{50}
\frac{ \bra{\Psi^0_{\mathrm G}} 
\widehat{e}_{{\mathbf p}, \te}^{\, +}\widehat{v}_{{\mathbf p}, \te}^{\vphantom{\, +}} 
\ket{\Psi^0_{\mathrm G}}}{\left \langle \Psi^0_{\mathrm G} 
| \Psi^0_{\mathrm G} \right \rangle}=
\Theta(E_{\mathrm F}-\overline{E}({\mathbf p}, \te))\; ,
\end{equation}
at zero temperature.
We will see below that it is actually 
possible to define operators $\widehat{e}^{\, +}_{{\mathbf p},t}$ 
and $\widehat{v}_{{\mathbf p},t}^{\vphantom{\, +}}$ which obey~(\ref{50}) 
in the whole Brillouin zone and not only around the Fermi surface. 
This implies that our variational approach 
does not capture the damping of quasi-particles. 

\subsubsection{Quasi-particles for a rigid Fermi-sea background}

First, we adopt the viewpoint of a fixed Fermi-sea background,
i.e., we assume that a quasi-particle is added to the
$N$-particle system whose variational parameters have been fixed by
the minimization of the 
energy expression~(\ref{43}), or equivalently, by the conditions~(\ref{222}).
This leads to the optimum one-particle
product state for the $N$-particle system
\begin{equation}
\label{60}
\ket{\overline{\Phi}_0}=
\prod_{{\mathbf k}, \re;
\overline{E}({\mathbf k}, \re)<E_{\rm F}}
\widehat{\overline{h}}{}\vphantom{\widehat{h}}^{\, +}_{{\mathbf p},\re}
\ket{\text{vacuum}} \; ,
\end{equation}
which, in infinite dimensions,
actually is the ground state of the effective one-particle Hamiltonian
\begin{equation}
\label{70}
\widehat{\overline{H}}{}^{\mathrm{eff}}=  
\sum_{{\mathbf k}, \re}\overline{E}({\mathbf k}, \re)
\widehat{\overline{h}}{}\vphantom{\widehat{h}}_{{\mathbf k}, \re }^{\, +}
\widehat{\overline{h}}{}\vphantom{\widehat{h}}_{{\mathbf k}, \re }^{\vphantom{\, +}}\;.
\end{equation}
The conditions~(\ref{222}) furthermore lead 
to optimum parameters~$\overline{m}_{\Gamma}$ 
and by these means define an optimum 
correlation operator~$\widehat{\overline{P}}_{\mathrm G}$. 
Using the one-particle operators 
$\widehat{\overline{h}}{}\vphantom{\widehat{h}}_{{\mathbf p},\te}^{\, +}$ 
and $\widehat{\overline{h}}{}\vphantom{\widehat{h}}_{{\mathbf p},\te}^{\vphantom{\, +}}$ we can now identify
\begin{subequations}
\label{80}
\begin{eqnarray}
 \label{eqn:QT-ER}
 \widehat{e}_{{\mathbf p}, \te}^{\, +}& := &
\widehat{\overline{P}}_{\mathrm G} 
\widehat{\overline{h}}{}\vphantom{\widehat{h}}_{{\mathbf p},\te}^{\, +} 
(\widehat{\overline{P}}_{\mathrm G})^{-1} \;,\\
 \label{eqn:QT-VER}
  \widehat{v}_{{\mathbf p}, \te}^{\vphantom{\, +}} 
&:=&\widehat{\overline{P}}_{\mathrm G}
 \widehat{\overline{h}}{}\vphantom{\widehat{h}}_{{\mathbf p},\te}^{\vphantom{\, +}} 
(\widehat{\overline{P}}_{\mathrm G})^{-1}
\end{eqnarray}\end{subequations}%
as those operators which obey the quasi-particle condition~(\ref{50}). 
Note that the inverse operator $(\widehat{\overline{P}}_{\mathrm G})^{-1}$ 
in~(\ref{80}) is well defined 
since we expect all parameters $\overline{m}_{\Gamma}$ 
to be finite in Fermi-liquid systems.   

Adding/removing a quasi-particle to/from
the ground state generates the excited states
\begin{subequations}
\label{90}
\begin{eqnarray}
| \overline{\Psi}_{\mathrm G}^{({{\mathbf p}}, \te)} \rangle_{\rm qp} &=&    
\widehat{e}_{{\mathbf p}, \te}^{\, +} | 
\overline{\Psi}^0_{\mathrm G} \rangle =   
\widehat{\overline{P}}_{\mathrm G} 
\widehat{\overline{h}}{}\vphantom{\widehat{h}}_{{\mathbf p}, \te}^{\, +}
| \overline{\Phi}_0 \rangle\;, \\
| \overline{\Psi}_{\mathrm G}^{({\mathbf p}, \te)} \rangle_{\rm qh} &=&    
\widehat{v}_{{\mathbf p}, \te}^{\vphantom{\, +}} 
| \overline{\Psi}^0_{\mathrm G} \rangle 
=\widehat{\overline{P}}_{\mathrm G} 
\widehat{\overline{h}}{}\vphantom{\widehat{h}}_{{\mathbf p}, \te}^{\vphantom{\, +}}  
| \overline{\Phi}_0 \rangle
\end{eqnarray}\end{subequations}%
with fixed Fermi-liquid background.
As described in Sect.~\ref{sec2}, these equations 
constitute an explicit example for Landau's ideas.
The Gutzwiller correlator~$\widehat{\overline{P}}_{\rm G}$ in~(\ref{30})
adiabatically transforms Fermi-gas eigenstates $|\Phi\rangle$
into (approximate) eigenstates of the Fermi liquid. 

The energy of quasi-particles or quasi-holes is defined as
\begin{equation}
\label{100}
\overline{E}_{\rm qp}({\mathbf p},\te) = 
\pm (\overline{E}^{\rm var}_0({\mathbf p},\te) - E^{\rm var}_0) -\mu
\end{equation}
where
\begin{subequations}
\label{110global}
\begin{eqnarray}
E_0^{\rm var}&=&
\langle \widehat{H} \rangle_{\overline{\Psi}_{\rm G}^0}=(\ref{224}) \;,
\label{110}
\\
\overline{E}^{\rm var}_0({\mathbf p}, \te)
&=&\langle \widehat{H} 
\rangle_{\overline{\Psi}_{\mathrm G}^{({\mathbf p}, \te)}}\;.
\label{112}
\end{eqnarray}\end{subequations}%
The $\pm$-sign refers to quasi-particle or quasi-hole states, respectively.
We define the quasi-particle energy~(\ref{100}) in reference to the
(variational) chemical potential of the system, see~(\ref{114}).
Note that the energy in~(\ref{100}) is of order unity whereas
those in~(\ref{110global}) are of ${\cal O}(L)$.

The definition of the quasi-particle states~(\ref{90})
applies to systems of arbitrary spatial dimensions. However, in
finite dimensions, the one-particle 
operators~$\widehat{\overline{h}}{}\vphantom{\widehat{h}}_{{\mathbf p}, \te}$
in~(\ref{60}) and~(\ref{90})
cannot be derived from the diagonalization of the effective Hamiltonian~(\ref{170});
in this case, 
the operators~$\widehat{\overline{h}}{}\vphantom{\widehat{h}}_{{\mathbf p}, \te}$ 
must be determined by a minimization of the variational ground-state energy
with respect to the amplitudes $F_{{\mathbf k},\s,\re }^{\vphantom{\ast}}$
in~(\ref{180}).

\subsubsection{Quasi-particles with background relaxation}

When we add a particle to the $N$-particle system 
we may expect that the variational parameters will adjust to the
presence of the additional particle. 
Therefore, we may want to work with
\begin{equation}
\label{140}
| \Psi_{\mathrm G}^{({\mathbf p}, \te)} \rangle_{\rm qp} 
=
\widehat{P}_{\mathrm G} \widehat{h}_{{\mathbf p}, \te}^{\, +}   
| \Phi_0 \rangle  
\quad , \quad
| \Psi_{\mathrm G}^{({\mathbf p}, \te)} \rangle_{\rm qh} 
= \widehat{P}_{\mathrm G} \widehat{h}_{{\mathbf p}, \te}^{\vphantom{\, +}}   
| \Phi_0 \rangle  \; ,
\end{equation}
where the $N$-particle
Fermi sea $|\Phi_0\rangle$ is defined according to~(\ref{190}).
Note that the operators $\widehat{h}_{{\mathbf p},\te}^{\, +}$,
$\widehat{h}_{{\mathbf p},\te}^{\vphantom{\, +}}$ still depend
on the parameters $m_{\Gamma}$, $n_{\s}$, and $\lambda_{\s}$
for a system with $N\pm 1$ particles. 
Then, 
\begin{equation}
\label{120}
E_{\rm qp}({\mathbf p},\te) 
=\pm (E^{\rm var}_0({\mathbf p},\te) - E^{\rm var}_0)  -\mu
\end{equation} 
with
\begin{equation}
\label{130}
E^{\rm var}_0({\mathbf p},\te)
=\mathop{\rm Min}_{m_{\Gamma},n_{\s},\lambda_{\s},\Lambda} 
\Bigl[\langle \widehat{H}  
\rangle_{\Psi_{\mathrm G}^{({\mathbf p}, \te)}}
-L \Lambda(n\pm\frac{1}{L}-\sum_{\s}n_{\s})  \Bigr] 
\end{equation}
is the definition of the quasi-particle and quasi-hole
energy with background relaxation. 

One may wonder whether the two definitions~(\ref{100}) and~(\ref{120}) 
will lead to different results
for the energies of quasi-particles and quasi-holes.
Fortunately, this is not the case in the thermodynamic limit, i.e.,
\begin{equation}
\overline{E}_{\rm qp}({\mathbf p},\te) 
= E_{\rm qp}({\mathbf p},\te) +{\cal O}(1/L)\; ,
\end{equation}
as we will show explicitly in appendix~\ref{appa1}.
The addition/subtraction of one particle
leads to a change in the optimized variational parameters 
to order $(1/L)$, and, in principle, this could result
in a change of the variational energy $E_0^{\rm var}$
to order unity. 
However, this quantity is
extremal with respect to the variational parameters,
so that it changes only to order~$(1/L)$ for parameter variations
around their optimal values. Therefore, the change
of the quasi-particle energies due to the background relaxation  
vanishes in the thermodynamic limit.

\subsection{Quasi-particle dispersion}
\label{sec4.2}

In the following we focus on $E_{\rm qp}({\mathbf p},\te)$
because the evaluation of~(\ref{100}) is more involved.
The energy~(\ref{130}) is given by 
\begin{equation}
\label{260}
E^{\rm var}_0({\mathbf p},\te)=
\mathop{\rm Min}_{m_{\Gamma},n_{\s},\lambda_{\s},\Lambda}
\left[ \widetilde{E}^{({\mathbf p},\te)}_{\mathrm c} 
[m_{\Gamma},n_{\s},\lambda_{\s},\Lambda] \right]  
\end{equation}
where
{\arraycolsep=1.5pt\begin{eqnarray}
\widetilde{E}^{({\mathbf p},\te)}_{\mathrm c} 
[m_{\Gamma},n_{\s},\lambda_{\s},\Lambda]
&=&
E^{({\mathbf p},\te)}_{\rm SP}
+\sum_{\Gamma}E_{\Gamma} m_{\Gamma}-L\sum_{\s}\lambda_{\s} n_{\s}
\nonumber \\
&&
-L \Lambda(n\pm\frac{1}{L}-\sum_{\s}n_{\s}) 
\label{270}
\end{eqnarray}}%
and 
\begin{equation}
\label{280}
E^{({\mathbf p},\te)}_{\rm SP}=
\pm E({\mathbf p},\te)+\sum_{{\mathbf k}, \re}
E({\mathbf k}, \re) \Theta(E_{\rm F} - E({\mathbf k}, \re)) \;.
\end{equation}
The ($\pm$) sign in~(\ref{270}) and~(\ref{280}) 
correspond to a quasi-particle and quasi-hole state, respectively.

Adding or removing a particle changes the 
parameters $m_{\Gamma}$, $n_{\s}$, $\lambda_{\s}$, and $\Lambda$ 
and the energies $E({\mathbf k},\te)$ 
only by terms of the order $(1/L)$ compared 
to their values in the $N$-particle ground state,
\begin{subequations}
\label{290global}
\begin{eqnarray}
x_i&=&\overline{x}_i+\frac{\delta x_i}{L}\quad\text{with}\quad
x_i \in \{m_{\Gamma},n_{\s},\lambda_{\s},\Lambda\}\;,\nonumber \\
&& \label{290}
\\
E({\mathbf k},\te)&=&\overline{E}({\mathbf k},\te)
+\frac{\delta E({\mathbf k},\te)}{L}\;.
\end{eqnarray}\end{subequations}%
Thus we may expand~(\ref{270}) in terms of~$(1/L)$ up to order unity, 
\begin{eqnarray}
\widetilde{E}^{({\mathbf p},\te)}_{\mathrm{c}}
[m_{\Gamma},n_{\s},\lambda_{\s},\Lambda]
&=&
\widetilde{E}_{\mathrm{c}} 
[\overline{m}_{\Gamma},\overline{n}_{\s},\overline{\lambda}_{\s},
\overline{\Lambda}]
\nonumber \\[3pt]
&& 
\pm(\overline{E}({\mathbf p},\te)-\overline{\Lambda})
+\delta\widetilde{E}_{\rm c}
\label{300}
\\ 
\delta\widetilde{E}_{\rm c} &=&
\!\!\!\!\sum_{x_i \in \{m_{\Gamma},n_{\s},\lambda_{\s},\Lambda\}}
\left.\frac{\partial}{\partial x_i}\widetilde{E}_{\mathrm c}  
\right|_{x_i=\overline{x}_i}\frac{\delta x_i}{L}\, .
\nonumber \\
&& \label{301}
\end{eqnarray}
The sum in~(\ref{301}) vanishes according to~(\ref{222}).
Using~(\ref{246}) the quasi-particle dispersion~(\ref{120}) becomes
\begin{equation}
\label{310}
E_{\rm qp}({\mathbf p},\te)=\overline{E}({\mathbf p},\te)-E_{\rm F}\;.
\end{equation}
This result does not come as a surprise since 
the Fermi surfaces, as defined by the conditions
$\overline{E}({\mathbf p},\te)=E_{\rm F}$ 
and $E_{\rm qp}({\mathbf p},\te)=0$, must
coincide. In addition, eq.~(\ref{310}) shows that the quasi-particle
dispersion is given by the eigenvalues 
$\overline{E}({\mathbf p},\te)$ of the effective Hamiltonian~(\ref{70}) 
not only around the Fermi surface but in the whole Brillouin zone. 
Note that the variational kinetic energy,
\begin{equation}
\langle \hat{H}_1    \rangle_{\overline{\Psi}_{\rm G}^0}
= \langle \widehat{\overline{H}}{}^{\mathrm{eff}}
- L \sum_{\sigma}\overline{\lambda}_{\s} \hat{n}_{\s}
\rangle_{\overline{\Phi}_0}  \; ,
\end{equation}
is given by the expectation value of the effective Hamiltonian~(\ref{70})
only in the case of degenerate orbitals where $\overline{\lambda}_{\s}=0$.

There are two important differences 
between the effective Hamiltonian~(\ref{70}), or, equivalently,~(\ref{170}), 
and the bare one-particle Hamiltonian
$\widehat{H}_1$ in~(\ref{10}). First, the bands are
narrowed in $\widehat{\overline{H}}{}^{\mathrm{eff}}$  
because the Coulomb interaction
reduces the mobility of the electrons. Second, the fields $\lambda_{\s}$ which
were originally introduced as auxiliary Lagrange parameters act as
observable shifts of the energy bands, e.g., in terms of a magnetic
exchange splitting. Our  
detailed numerical investigations on Nickel~\cite{gutzjubel,wandlitz}
showed that both effects, i.e., band-narrowing and band-shifts, are relevant
for a proper description of quasi-particles in real materials.

\section{1/D corrections for the one-band Hubbard model}
\label{sec5}

The energy expression~(\ref{43}) for the wave function~(\ref{30}) 
is exact in the limit of infinite spatial dimensions~$D$. 
Therefore, its evaluation for real, finite-dimensional systems 
constitutes an additional approximation. 
For the one-band model it has been shown~\cite{florian} 
that $1/D$~corrections of ground-state properties 
are actually small in most cases of interest. 
An exception is the half-filled Hubbard model where, 
in infinite dimensions, the Gutzwiller theory predicts 
the so-called Brinkman-Rice transition where all electrons become
localized at some finite critical interaction strength~$U_{\rm BR}$.
This metal-insulator is known to be an artifact 
of the limit $D\to\infty$ because it is absent 
in all finite dimensions~\cite{PvD}. 
Consequently, $1/D$ corrections must become important 
in this special case.

\subsection{First order corrections: analytical results}
\label{sec5.1}

In the case of only one orbital per lattice site, 
the general Hamiltonian~(\ref{10}) reduces to
\begin{equation}
\label{400}
\hat{H}=\sum_{\mathbf k}\sum_{\s=\uparrow,\downarrow}
\varepsilon(\mathbf k)
\hat{c}_{\mathbf k,\s}^{+}
\hat{c}_{\mathbf k,\s}^{\vphantom{+}}
+ U\sum_i \hat{n}_{i\uparrow} \hat{n}_{i\downarrow}\;. 
\end{equation}
We consider a hyper-cubic lattice with only nearest-neighbor 
hopping-terms where the  bare dispersion in~(\ref{400}) is given by
\begin{equation}
\label{402}
\varepsilon({\mathbf k})=-\sqrt{\frac{2}{D}}\sum_{l=1}^{D}\cos(k_{l})\;. 
\end{equation}
The Gutzwiller wave function will be evaluated 
in its original form~\cite{florian}, i.e.,
the variational parameter~$\lambda_{(\uparrow\downarrow)}$ for 
the doubly occupied state $\ket{\Gamma}=\ket{\uparrow\downarrow}$ 
is replaced by the parameter~$g$. 
For the one-band model both definitions are equivalent. 

The variational ground-state energy of the Hamiltonian~(\ref{400}) 
in infinite dimension reads
\begin{equation}
\label{410}
E^\infty(g,n)= L \left [ q(g,n) \overline{\varepsilon}_0+ 
U \overline{d}(g,n) \right ] \;,
\end{equation}
where
\begin{equation}
\label{420}
  \overline{\varepsilon}_0:= 
\frac{1}{L} 
\sum_{{\mathbf k}, \sigma} n_{{\mathbf k}, \sigma}^0 
\varepsilon({\mathbf k})
\end{equation}
is the mean kinetic energy of the non-interacting system.
Here, the renormalization factors~$q(g,n)$ and the average 
double occupancy per lattice site~$\overline{d}(g,n)$ are given by
\begin{subequations}
\begin{eqnarray}
q(g,n)&=&\frac{4}{n(2-n)} \Big(\frac{n}{2} -
\overline{d}(g,n) \Big) \nonumber \\
&& \times \Big ( \sqrt{1-n+\overline{d}(g,n)}
+\sqrt{\overline{d}(g,n)} \Big)^2 \, , \\
\overline{d}(g,n)&=&\frac{n}{2} \frac{G+n-1}{G+1}
\end{eqnarray}\end{subequations}%
with
\begin{equation}
\label{425}
  G=\sqrt{1+n(2-n)(g^2-1)} \;.
\end{equation}
The momentum distribution of the non-interacting system,
\begin{equation}
n_{{\mathbf k}, \sigma}^0= 
\Theta(E_{\mathrm F}-\varepsilon(\mathbf k))\;,
\end{equation}
and the electron density,
\begin{equation}
\frac{1}{L} \sum_{{\mathbf k},\sigma}n^{0}_{{\mathbf k}, \sigma}=n \; ,
\end{equation}
determine the Fermi energy $E_{\rm F}$.

We set up the $1/D$~expansion of a function $A(g,n)$ in the form
\begin{equation}
A(g,n) = A^{\infty}(g,n) + \frac{1}{D} A^{(1)}(g,n) +\ldots \; .
\end{equation}
Then, the first-order correction of the ground-state energy 
reads 
\begin{equation}
\label{440}
E^{(1)}(g,n)=L \left  [ t^{(1)}(g,n)+  U \overline{d}^{(1)}(g,n) \right ]\;,   
\end{equation}
where the corrections to the average kinetic energy and the double occupancy 
can be written as  
\begin{eqnarray}
\label{eqn:E-Kin-1d}
t^{(1)}(g,n)&=& \frac{1}{L}
\sum_{{\mathbf k},\sigma} n_{{\mathbf k},\sigma}^{(1)} 
\varepsilon_{\sigma}(\mathbf k)\;, \\
n_{{\mathbf k},\sigma}^{(1)}(g,n) &=&  
f(g,n)\left [ \frac{(n-1)(G-1)}{n(2-n)G}
\overline{\varepsilon}_0 + \varepsilon_{\sigma}({\mathbf k}) \right ] 
(\overline{\varepsilon}_0)^3
\nonumber \\
 &&\times [n(G+1-n)+2(1-n)(G-1)n_{{\mathbf k}, \sigma}^0] 
\nonumber \;,\\
&&   
\end{eqnarray} 
and
\begin{equation}
\label{460}
\overline{d}^{(1)}(g,n)= h(g,n) (\overline{\varepsilon}_0)^4 \;.
\end{equation}
Here, we introduced the factors
\begin{subequations}
\label{eqn:G-f}
\begin{eqnarray}
 f(g,n)&=& -\left( \frac{1}{1+g} \right )^2 
\left( \frac{G-1}{G+1} \right)^2 
\left(\frac{1}{n(2-n)} \right)^3\;,\nonumber\\
&& \\
       h(g,n)&=&
\frac{(G+1-n)(G+n-1)(G-1)}{2G(G+1)^3n^2(2-n)^2}\;.
\end{eqnarray}\end{subequations}%
The total ground-state energy to first order in~$1/D$, 
\begin{equation}
  \label{500}
   E(g,n)=E^{\infty}(g,n)+\frac{1}{D} E^{(1)}(g,n)\;, 
\end{equation}
has to be minimized with respect to~$g$. 
However, the optimum value $\overline{g}$ 
of this minimization differs from the respective 
value $\overline{g}^{\infty}$ 
in infinite dimensions only by terms of the order~$1/D$,
\begin{equation}
  \overline{g}=\overline{g}^{\infty}+\frac{1}{D} \overline{g}^{(1)} \;.
\end{equation}
Therefore, we can expand the optimum ground-state energy 
in terms of $1/D$ as
\begin{subequations}
\begin{eqnarray}
     E(\overline{g},n)&=&
E^{\infty}(\overline{g}^{\infty}+\frac{1}{D} \overline{g}^{(1)},n)
\nonumber \\
&& 
+\frac{1}{D} E^{(1)}
(\overline{g}^{\infty}+\frac{1}{D} \overline{g}^{(1)},n)
\label{530} \\
&\approx& E^{\infty}(\overline{g}^{\infty},n) 
+\frac{\overline{g}^{(1)}}{D} \left.
\frac{\partial E^{\infty} (g,n)}{\partial g}
\right|_{g=\overline{g}^{\infty}} 
\nonumber \\
&& 
+ \frac{1}{D} E^{(1)}(\overline{g}^{\infty},n) \;,
\label{532}
\end{eqnarray}\end{subequations}%
which leads to
\begin{equation}
\label{540}
   E(\overline{g},n)\approx E^{\infty}(\overline{g}^{\infty},n) 
+ \frac{1}{D} E^{(1)}(\overline{g}^{\infty},n)=
E(\overline{g}^{\infty},n)
  \end{equation}
because the derivative in~(\ref{532}) vanishes. From~(\ref{540}) 
we see that the optimum ground-state energy is determined 
by the optimum parameter $\overline{g}^{\infty}$ in infinite dimensions
and no minimization of the total energy~(\ref{500}) is required.

In order to determine the quasi-particle dispersion
as defined in~(\ref{100}), we evaluate~(\ref{110}) 
and~(\ref{112}) to order~$1/D$. The energy~(\ref{110}) 
is given by~(\ref{540}). The expression~(\ref{540}) 
also yields the energy~(\ref{112}) when we perform the replacements 
\begin{subequations}
\begin{eqnarray}
n&\to&n \pm \frac{1}{L}\\
n_{{\mathbf k}, \sigma}^0 &\to& n_{{\mathbf k}, \sigma}^0 
\pm \delta_{{\mathbf k},{\mathbf p} }\delta_{\s,\tau}
\end{eqnarray}\end{subequations}%
for a quasi-particle state ($+$~sign) 
or quasi-hole state ($-$~sign) 
with momentum ${\mathbf p}$ and spin~$\tau$. 
A straightforward expansion of~(\ref{112}) in terms of~$1/L$ 
leads to the quasi-particle energy 
\begin{equation}
\overline{E}_{\rm qp}({\mathbf p},\tau)=
\overline{E}^{\infty}_{\rm qp}({\mathbf p},\tau)
+\frac{1}{D} \overline{E}^{(1)}_{\rm qp}({\mathbf p},\tau)\;.
\end{equation}
Here, we recover the quasi-particle dispersion in infinite dimensions, 
\begin{equation}
\overline{E}^{\infty}_{\rm qp}({\mathbf p},\tau)=
\pm q(\overline{g}^{\infty},n)(\varepsilon({\mathbf p})-E_{\rm F})\;, 
\end{equation}
as already derived in Sec.~\ref{sec4}. 
The first-order correction reads
\begin{equation}
\overline{E}^{(1)}_{\rm qp}({\mathbf p},\tau)=\pm \big[
\varepsilon({\mathbf p}) - E_{\rm F} \big] 
\widetilde{E} \big[ \varepsilon({\mathbf p}) \big]
\end{equation}
with
{\arraycolsep=1pt\begin{eqnarray}
\label{580}
    \widetilde{E} \big[ \varepsilon({\mathbf p}) \big]
&=&
f(\overline{g}^{\infty},n) (\overline{\varepsilon}_0)^2 
\Big\{-10 
\frac{(n-1)^2(\overline{G}^{\infty}-1)^2}{n(2-n)\overline{G}^{\infty}}
(\overline{\varepsilon}_0)^2 \nonumber \\
&& + 6n(\overline{G}^{\infty}+1-n)  
+  6(1-n)(\overline{G}^{\infty}-1) \overline{\varepsilon_0^2} 
\nonumber \\
&&
+ \left ( \varepsilon({\mathbf p})
+  E_{\rm F} \right) 2(1-n)(\overline{G}^{\infty}-1) 
\overline{\varepsilon}_0 \Big \} 
\\ 
&& +4U h(\overline{g}^{\infty},n) (\overline{\varepsilon}_0)^3\;.
\nonumber
\end{eqnarray}}%
The quantity $\overline{G}^{\infty}$ is given by~(\ref{425}), 
evaluated at $g=\overline{g}^{\infty}$, 
and $\overline{\varepsilon_0^2}$ is defined by
\begin{equation}
\label{eqn:zweites-Moment}
\overline{\varepsilon_0^2}=\frac{1}{L} 
\sum_{{\mathbf k},\sigma} 
n_{{\mathbf k}, \sigma}^0 \varepsilon({\mathbf k})^2\;.  
\end{equation}
Note that in deriving~(\ref{580}) we have used the relations
\begin{equation}
  \sum_{{\mathbf k}} \varepsilon({\mathbf k})=0 \quad , 
\quad \frac{1}{L} \sum_{\mathbf k} [\varepsilon({\mathbf k})]^2=1  \; ,
\end{equation}
which hold for the dispersion relation~(\ref{402}).  

\subsection{First-order corrections: numerical results}
\label{sec5.2}

We are interested in the relative size of the $1/D$~corrections 
compared to the result in $D=\infty$ dimensions. 
For this purpose we introduce
\begin{equation}
\label{eqn:mass}
m({\mathbf p})=
\frac{\overline{E}_{\rm qp}({\mathbf p}, \tau)}{
\overline{E}_{\rm qp}^{\infty}({\mathbf p}, \tau)}-1
\end{equation}
as a measure for the deviations from the result in
infinite dimensions.

\begin{figure}
\includegraphics[width=6.6cm,angle=-90]{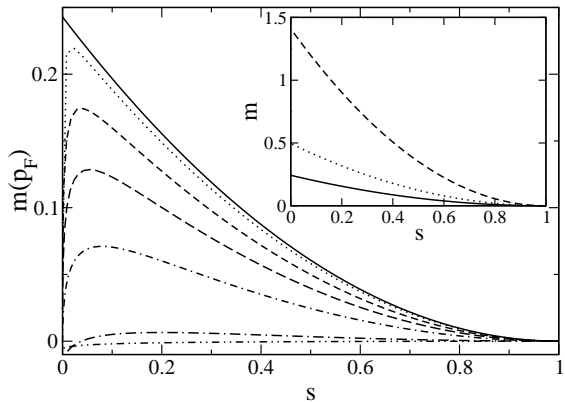}
\caption{Renormalization factor~$m({\mathbf p}_{\rm F})$
for the quasi-particle
dispersion at the Fermi energy as a function of $s=4\overline{d}/n^2$
for band fillings 
$n=1, 0.99, 0.95, 0.9, 0.8, 0.5, 0.2$ (from top to bottom at $s=0.2$)
in three dimensions.
Inset: special case of half band-filling for dimensions $D=1$ (dashed line),
$D=2$ (dotted line), and $D=3$ (full line).}
\label{fig1new} 
\end{figure}

In the half-filled case, $n=1$, the ratio $m({\mathbf p})$ 
is independent of the wave vector~${\mathbf p}$,
\begin{equation}
 m=m({\mathbf p})=\frac{1}{2D} 
\frac{\overline{g}^{\infty}-1}{\overline{g}^{\infty}+1} \overline{\varepsilon}_0^2 
\left [ \overline{\varepsilon}_0 U 
-\frac{3(\overline{g}^{\infty}-1)}{\overline{g}^{\infty}+1}\right ]\;.
\end{equation}
The inset of Fig.~\ref{fig1new} shows 
$m=m({\mathbf p})$ as a function of $s=4\overline{d}/n^2$ 
for spatial dimensions $D=1,2,3$. 
Here, $0\leq s\leq 1$ provides a measure for the 
correlation strength in the system. The value $s=1$ 
corresponds to $U=0$ and 
$s=0$ is realized at the Brinkman-Rice transition, $U=U_{\rm BR}=
8|\overline{\varepsilon}_0|$.
As seen from the inset of Fig.~\ref{fig1new},
$1/D$~corrections are not negligible over a wide range 
of interactions, especially in one dimension. 
In three dimensions, these corrections are much smaller but still 
about~$25\%$ close to the Brinkman-Rice transition.

Since this transition is spurious in finite dimensions,
$1/D$~corrections have to be large in the half-filled Hubbard model. 
For an application of our method to real Fermi-liquid systems 
it is more reasonable to study cases of non-integer band filling. 
Fig.~\ref{fig1new} shows the ratio~$m({\mathbf p}_{\rm F})$
at the Fermi surface for different band fillings
$n=1, 0.99, 0.95, 0.9, 0.8, 0.5, 0.2$ in three dimensions.  
The respective results for band-fillings
$n^{\prime}=2-n$ follow identically due to particle-hole symmetry.
As expected, the corrections in
Fig.~\ref{fig1new} become much smaller away from integer filling. 

\begin{figure}
\includegraphics[width=6.6cm,angle=-90]{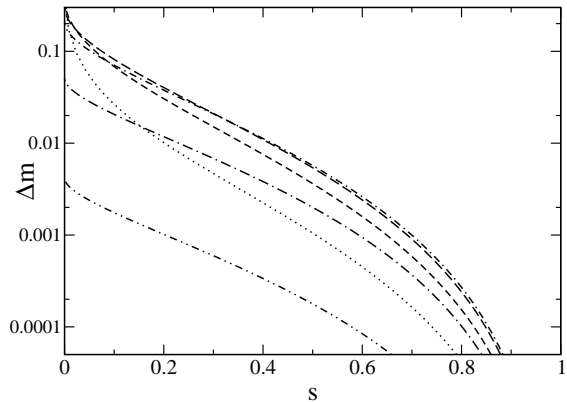}
\caption{Maximal width of the renormalization factor~$\Delta m$ 
for the quasi-particle
dispersion as a function of $s=4\overline{d}/n^2$ for band fillings
$n=1, 0.99, 0.95, 0.9, 0.8, 0.5, 0.2$ in three dimensions,
shown on a logarithmic scale; notation 
as in Fig.~\protect\ref{fig1new}.}
\label{fig2new} 
\end{figure}

The data in Fig.~\ref{fig1new} show $m({\mathbf p}_{\rm F})$ 
at the Fermi energy. However, for $n\neq 1$, there also is a 
momentum dependence of $m({\mathbf p})$ which can become significant
close to half band filling. In Fig.~\ref{fig2new} 
we show the width of first order contributions,
\begin{equation}
\Delta m=\mathop{\rm Max}_{{\mathbf p}} 
|m({\mathbf p})-m({\mathbf p}_{\rm F})|\; ,
\end{equation}
on a logarithmic scale for the same band fillings as 
in Fig.~\ref{fig1new}. Although ~$\Delta m$ strictly vanishes 
for $n=1$  we see from Fig.~\ref{fig2new} that~$\Delta m$ 
can become relatively large for $n\lesssim 1$. 
This means that around the half-filled case we find $1/D$~corrections 
which strongly depend on the wave vector. 

As long as $\Delta m\ll  m({\mathbf p}_{\rm F})$, 
a finite value of~$m({\mathbf p})$
amounts to a rescaling of
the overall band width.
When we apply our theory to real materials~\cite{gutzjubel,wandlitz}
the band width is basically controlled by the Racah-parameter~$A$ 
which we adjust to fit the experimental band width. 
Therefore, $1/D$~corrections without a significant 
momentum dependence will not modify the band structure 
in our variational approach. 

As shown in Figs.~\ref{fig1new},~\ref{fig2new} 
the results for $D\to \infty$ become questionable 
only close to integer filling and for very strong correlations.
Therefore, we have reasons to believe that
the quasi-particle dispersions as calculated in $D=\infty$ in
Sect.~\ref{sec4} provide a good starting point for a sensible
comparison with experimental data. 
The good agreement between experiments 
and our theoretical results on nickel~\cite{gutzjubel,wandlitz} 
supports such an optimistic point of view.

\section{Conclusions}
\label{sec6}

In this work we used Landau's Fermi-liquid picture to define
quasi-particle excitations in terms of Gutzwiller-correlated 
wave functions. Starting from the optimum variational ground state of a 
general multi-band Hubbard-model 
we identified operators which describe the creation and 
annihilation of quasi-particles in this state. 
We calculated the quasi-particle dispersion
analytically in the limit of infinite dimensions.
Our variational states provide an illustrative 
example for Landau quasi-particles. They 
are also suitable for numerical investigations, e.g.,
with variational Monte-Carlo techniques.

We gave two definitions of quasi-particle operators,
with and without a relaxation of the Fermi-sea background.
It turns out that it is more convenient to allow
a (small) change of the variational parameters of the Fermi-sea
background in the presence of the quasi-particles.
We showed that both cases lead to the same result 
for the quasi-particle dispersion.
This absence of a orthogonality catastrophe is characteristic for
Fermi liquids. 

Our results confirm our earlier calculations in which the 
quasi-particle dispersion had been extracted phenomenologically 
from an effective one-particle Hamiltonian~\cite{PRB98,gutzjubel,wandlitz}. 
In contrast to density-functional theory,
our quasi-particle dispersions correspond to
mathematically well-defined (variational) states
in realistic multi-band Hubbard models. 
In general, our quasi-particle bands are 
narrower than the DFT bands because of the hopping-reduction factors $q_{\s}$
in~(\ref{43}).
Moreover, as seen in~(\ref{170}), the Gutzwiller theory has the flexibility
for the adjustment of the orbital energies through the parameters $\lambda_{\s}$
so that the DFT bands are shifted and mixed into the Landau-Gutzwiller
quasi-particle bands.

Our derivation of the Gutzwiller theory uses approximations 
which become exact in the limit of infinite spatial dimensions,
$D\to\infty$. 
For this reason, we calculated first-order corrections in $1/D$
for the quasi-particle dispersion of the one-band Hubbard model. 
Apart from the special case close to half band-filling, 
these corrections were found to be relatively small.
Consequently,
the quasi-particle bands as derived in $D=\infty$ for
multi-band Hubbard models contain the essential information
of the Gutzwiller states in three dimensions, and are thus suitable
for a meaningful comparison with
real, three-dimensional Fermi liquids.   

\begin{acknowledgments}
J.B.~gratefully acknowledges support by the Deutsche Forschungsgemeinschaft.
We thank W.~Weber for helpful discussions and his critical 
reading of the manuscript.
\end{acknowledgments}

\begin{appendix}
\section{Quasi-particle dispersion for a rigid Fermi-sea background}
\label{appa1}

In this appendix we evaluate the quasi-particle dispersion~(\ref{100}) 
and thereby prove that it is identical to the energy~(\ref{310}). 

The variational ground state $\ket{\overline{\Psi}^0_{\rm G}}$ in~(\ref{110}) 
is given as 
\begin{equation}
\label{a10}
\ket{\overline{\Psi}^0_{\rm G}}=
\widehat{\overline{P}}_{\mathrm G} \ket{\overline{\Phi}_0},
\end{equation}
where 
$\ket{\overline{\Phi}_0}$ is the state~(\ref{190}) 
evaluated for the optimum values 
$\overline{m}_{\Gamma}$, $\overline{n}_{\s}$, $\overline{\lambda}_{\s}$,
and $\overline{\Lambda}$. 
The variational ground-state energy~(\ref{110}) therefore reads
\begin{eqnarray}
E^{\rm var}_0 
&=&\sum_{\s,\s^\prime}\sum_{\mathbf k}
S_{\s,\s^\prime}(\mathbf k;\overline{n}_{\s},\overline{m}_{\Gamma }) 
\langle \widehat{c}_{\mathbf k;\s}^{\, +}
\widehat{c}_{\mathbf k;\s'}^{\vphantom{\, +}} \rangle_{\overline{\Phi}_0}
\nonumber \\
&& 
+L\sum_{\Gamma }E_{\Gamma}\overline{m}_{\Gamma }\;. 
\label{a12}
\end{eqnarray}
Here, we made it explicit that the numbers 
$S_{\s,\s^\prime}({\mathbf k})$ in~(\ref{43}) 
depend on $n_{\s}$ and $m_{\Gamma }$. 
The state~(\ref{90}), which determines the expectation value~(\ref{112})  
can be written as 
\begin{equation}
\label{a20}
| \overline{\Psi}_{\mathrm G}^{({\mathbf p}, \te)} \rangle
=\widehat{\overline{P}}_{\mathrm G} | 
\overline{\Phi}^{({\mathbf p}, \te)} \rangle \; ,
\end{equation}
where 
\begin{equation}
\label{a22}
\ket{\overline{\Phi}^{({\mathbf p}, \te)}} 
=\widehat{\overline{h}}{}\vphantom{\widehat{h}}_{{\mathbf p}, \te}^{\, (+)}\ket{\overline{\Phi}_0}\;.
\end{equation}
The densities $n_{\s}$ and the parameters $m_{\Gamma}$ 
for the state \ket{\overline{\Phi}^{({\mathbf p}, \te)}} 
differ from those of the 
$N$-particle ground state only by terms of the order~$1/L$,
\begin{subequations}
\label{a30}
\begin{eqnarray}
\label{a32}
n_{\s}&=&\overline{n}_{\s}+\frac{1}{L}\delta n_{\s}\;,\\
m_{\Gamma}&=&\overline{m}_{\Gamma }
+\frac{1}{L}\delta m_{\Gamma}\;,
\end{eqnarray}
where, for example,
\begin{equation}
\label{a40}
\delta n_{\s}=\pm|\overline{F}_{{\mathbf p},\s,\te}^{\vphantom{\ast}}|^2\;.
\end{equation}\end{subequations}%
Here, the signs $\pm$ refer to a quasi-particle or quasi-hole state, 
respectively. Using~(\ref{a22})--(\ref{a30})  
we can write the energy~(\ref{112}) as
\begin{eqnarray}
\overline{E}^{\rm var}_0({\mathbf p}, \te) &=&
\sum_{\s,\s^\prime}\sum_{\mathbf k}
S_{\s,\s^\prime}(\mathbf k;\overline{n}_{\s}
+\frac{1}{L}\delta n_{\s},\overline{m}_{\Gamma }
+\frac{1}{L}\delta m_{\Gamma}) \nonumber \\
&& \hphantom{\sum_{\s,\s^\prime}\sum_{\mathbf k}}
\times \langle \widehat{c}_{\mathbf k;\s}^{\, +}
\widehat{c}_{\mathbf k;\s'}^{\vphantom{\, +}} 
\rangle_{\overline{\Phi}^{({\mathbf p}, \te)}} \label{a50}
\\
&& +L\sum_{\Gamma }E_{\Gamma}(\overline{m}_{\Gamma }+
\frac{1}{L}\delta m_{\Gamma}  )\;. 
\nonumber
\end{eqnarray}
For the expectation value in~(\ref{a50}) we find 
\begin{eqnarray}
 \langle \widehat{c}_{\mathbf k;\s}^{\, +}
\widehat{c}_{\mathbf k;\s'}^{\vphantom{\, +}}
\rangle_{\overline{\Phi}^{({\mathbf p}, \te)}}&=& 
\langle \widehat{c}_{\mathbf k;\s}^{\, +}
\widehat{c}_{\mathbf k;\s'}^{\vphantom{\, +}}\rangle_{\overline{\Phi}_0}
\pm \delta_{\mathbf k ,{\mathbf p}  }    
\overline{F}_{{\mathbf p},\s,\te}^{\ast} 
\overline{F}_{{\mathbf p},\s^\prime,\te}^{\vphantom{\ast}} 
\nonumber \;.\\
&& \label{a60}
\end{eqnarray}
An expansion of~(\ref{a50}) in terms of $1/L$ up to
and including terms of order unity leads to
\begin{eqnarray}
\overline{E}^{\rm var}_0({\mathbf p}, \te) 
&=&
E^{\rm var}_0\pm \sum_{\s,\s^\prime}
S_{\s,\s^\prime}({\mathbf p};\overline{n}_{\s},\overline{m}_{\Gamma })
\overline{F}_{{\mathbf p},\s,\te}^{\ast} 
\overline{F}_{{\mathbf p},\s^\prime,\te}^{\vphantom{\ast}}  \nonumber
\\ 
&& +\sum_{\gamma}\delta n_{\gamma}\frac{1}{L}
\sum_{\s,\s^\prime}\sum_{\mathbf k}
\langle \widehat{c}_{\mathbf k;\s}^{\, +}
\widehat{c}_{\mathbf k;\s'}^{\vphantom{\, +}} 
\rangle_{\overline{\Phi}_0}   \nonumber \\
&& \hphantom{+\sum_{\gamma}   }
\times 
\left.\frac{\partial}{\partial n_{\gamma}} 
S_{\s,\s^\prime}(\mathbf k;n_{\s},\overline{m}_{\Gamma })
\right|_{n_{\s}=\overline{n}_{\s}}
\label{a70}
\\ 
&&  +\sum_{\Gamma^{\prime}}
\delta m_{\Gamma^{\prime}}
\Bigl[E_{\Gamma^{\prime}}
+\frac{1}{L}\sum_{\s,\s^\prime}
\sum_{\mathbf k}
\langle \widehat{c}_{\mathbf k;\s}^{\, +}
\widehat{c}_{\mathbf k;\s'}^{\vphantom{\, +}} 
\rangle_{\overline{\Phi}_0}  
\nonumber \\
&& \hphantom{  +\sum_{\Gamma^{\prime}} }
\times 
\left. \frac{\partial}{\partial m_{\Gamma^{\prime}}}
S_{\s,\s^\prime}(\mathbf k;\overline{n}_{\s},m_{\Gamma })
\right|_{ m_{\Gamma }= \overline{m}_{\Gamma }}\Bigr]\;.
\nonumber
\end{eqnarray}
With the help of equations~(\ref{169}) and~(\ref{170}) we find
\begin{subequations}
\begin{eqnarray}
\label{a80}
\frac{\partial}{\partial n_{\gamma}}E^{\rm SP}
&=&\left \langle \frac{\partial}{\partial n_{\gamma}}\widehat{H}^{\rm eff} 
\right  \rangle_{ \overline{\Phi}_0}\\
\label{a82}
&=&\sum_{\s,\s^\prime}
\sum_{\mathbf k}
\langle \widehat{c}_{\mathbf k;\s}^{\, +}
\widehat{c}_{\mathbf k;\s'}^{\vphantom{\, +}}
\rangle_{\overline{\Phi}} \nonumber \\
&& \hphantom{\sum_{\s,\s^\prime}}
\times \left.\frac{\partial}{\partial n_{\gamma}} 
S_{\s,\s^\prime}(\mathbf k;n_{\s},\overline{m}_{\Gamma })
\right|_{n_{\s}=\overline{n}_{\s}}
\\ 
\label{a84}
&=&L(\overline{\lambda}_{\gamma}-\overline{\Lambda})
\end{eqnarray}\end{subequations}%
where the third line~(\ref{a84}) follows from~(\ref{220}) 
and~(\ref{222}). In the same way we can show that
{\arraycolsep=0.5 pt\begin{eqnarray}
\sum_{\s,\s^{\prime}}\sum_{\mathbf k}
\langle \widehat{c}_{\mathbf k;\s}^{\, +}
\widehat{c}_{\mathbf k;\s'}^{\vphantom{\, +}} \rangle_{\overline{\Phi}_0}  
&&\left.\frac{\partial}{\partial m_{\Gamma^{\prime}}} 
S_{\s,\s^\prime}(\mathbf k;\overline{n}_{\s},m_{\Gamma })
\right|_{ m_{\Gamma }= \overline{m}_{\Gamma }}=
\nonumber \\
&&-L\cdot E_{\Gamma^{\prime}}\;.
\label{a100}
\end{eqnarray}}%
Therefore, the energy difference in~(\ref{100}) becomes
{\arraycolsep=1.5pt\begin{eqnarray}
\overline{E}^{\rm var}_0({\mathbf p}, \te)-E^{\rm var}_0
&=&\pm\sum_{\s,\s^\prime}
\Bigl(S_{\s,\s^\prime}({\mathbf p};\overline{n}_{\s},\overline{m}_{\Gamma })
\overline{F}_{{\mathbf p},\s,\te}^{\ast} 
\overline{F}_{{\mathbf p},\s^\prime,\te}^{\vphantom{\ast}} 
\nonumber \\
&&
+\delta_{\s,\s^\prime}(\overline{\lambda}_{\s}-\overline{\Lambda})|
\overline{F}_{{\mathbf p},\s,\te}^{\vphantom{\ast}}|^2 \Bigr)\;.
\label{a110}
\end{eqnarray}}%
The first two terms in this expression just give the 
eigenvalues $\overline{E}({\mathbf p},\te)$ 
of the effective Hamiltonian~(\ref{70}),
which leads to
\begin{equation}
\overline{E}^{\rm var}_0({\mathbf p}, \te)-
E^{\rm var}_0=\pm(\overline{E}({\mathbf p},\te)-\overline{\Lambda})\;.
\end{equation}
Thus, by use of~(\ref{248}), the quasi-particle dispersion~(\ref{112}) 
finally becomes
\begin{equation}
\overline{E}_{\rm qp}({\mathbf p},\te)=
\overline{E}({\mathbf p},\te)-E_{\rm F}\;,
\end{equation}
in agreement with our 
result for $E_{\rm qp}({\mathbf p},\te)$
as derived in Sec.~\ref{sec4.2}.
\end{appendix}

\end{document}